\documentclass{PoS}

\newcommand{\aj}{AJ}
\newcommand{\apj}{ApJ}
\newcommand{\apjl}{ApJL}
\newcommand{\apjs}{ApJS}

\newcommand{\mnras}{MNRAS}

\newcommand{\mpifr}{Max-Planck-Institut f\"ur Radioastronomie (MPIfR), Germany}

\newcommand{\etal}{et. al.}

\title{Real time cosmology - A direct measure of the expansion rate of
  the Universe}

\ShortTitle{Real time cosmology - A direct measure of the expansion rate of
  the Universe}

\author{\speaker{Hans-Rainer Kl\"ockner}\\
        \mpifr \\
        E-mail: \email{hrk@mpifr-bonn.mpg.de}}

\author{Danail Obreschkow\\
  The University of Western Australia, ICRAR, Australia\\
    E-mail: \email{danail.obreschkow@icrar.org}}

\author{Carlos Martins\\
Centro de Astrof\'{\i}sica da Universidade do Porto, Portugal\\
    E-mail: \email{Carlos.Martins@astro.up.pt}}

\author{Alvise Raccanelli\\
  Department of Physics \& Astronomy, Johns Hopkins University, Caltech, \& JPL, USA\\
    E-mail: \email{alvise@caltech.edu}}

\author{David Champion\\
        \mpifr \\
    E-mail: \email{champion@mpifr-bonn.mpg.de}}

\author{Alan Roy\\
       \mpifr \\
    E-mail: \email{aroy@mpifr-bonn.mpg.de}}

\author{Andrei Lobanov\\
       \mpifr \\
    E-mail: \email{alobanov@mpifr-bonn.mpg.de}}

\author{Jan Wagner\\
 \mpifr\  \& KASI
  KVN, Korea\\
    E-mail: \email{jwagner@mpifr-bonn.mpg.de}, \email{jan.wagner@iki.fi}}

\author{Reinhard Keller\\
       \mpifr \\
    E-mail: \email{rkeller@mpifr-bonn.mpg.de}}

  \abstract{In recent years cosmology has undergone a revolution, with
    precise measurements of the microwave background radiation, large
    galaxy redshift surveys, and the discovery of the recent
    accelerated expansion of the Universe using observations of
    distant supernovae. All these ground-breaking observations have
    boosted our understanding of the Cosmos and its evolution. Because
    of this detailed understanding, more detailed tests of
    cosmological models require unprecedented precision that is only
    available with the next generation of astronomical
    observatories. Radio observations in particular will be able to
    access more independent modes than optical, infrared or X-ray
    facility and will show very different systematics compared to
    these other wavebands.

    \smallskip

    In this light, the SKA enables us to do an ultimate test in
    cosmology by measuring the expansion rate of the Universe in real
    time. This can be done by a rather simple experiment of observing
    the neutral hydrogen (HI) signal of galaxies at two different
    epochs. The signal will encounter a change in frequency imprinted as
    the Universe expands over time and thus monitoring the drift in
    frequencies will provide a real time measure of the cosmic
    acceleration. Over a period of 12~years one would expected a
    frequency shift of the order of 0.1~Hz assuming a standard
    $\Lambda$CDM cosmology. However, monitoring such changes would
    require some modifications to the current baseline design of the
    SKA. In particular, the design needs to be adapted to achieve
    higher spectral resolution, at least within sub-bands (strong
    requirement), and to allow for a well monitored distribution of
    the local oscillator signal, preferably at milli-Hz accuracy over
    a period of 12~years (weaker requirement, which could be
    circumvented by pulsar observations). Based on the sensitivity
    estimates of the SKA and the number counts of the expected HI
    galaxies, it is shown that the number counts are sufficiently high
    to compensate for the observational uncertainties of the
    measurements and hence allow a statistical detection of the
    frequency shift. In addition, depending on the observational
    setup, it is shown that the evolution of the frequency shift in
    redshift space can be estimated to a precision of a percent.

    \smallskip

    Although technically challenging, real time cosmology will be possible
    with the SKA due to its superior survey capabilities and the
    ability to measure Milk-Way type galaxies up to redshifts of unity.
    The direct measurement of the frequency shift and hence the cosmic
    acceleration can provide a model independent confirmation of dark
    energy. At highest precision it can distinguish between some
    competing cosmological models and combined with probes at other
    wavelength can break degeneracies and improving the figure of
    merit of cosmological parameters.
}

\FullConference{
Advancing Astrophysics with the Square Kilometre Array\\
June 8-13, 2014\\
Giardini Naxos, Italy}

\begin{document}

\section{Introduction}

The Big Bang concept of our Universe is well established as ``the
standard model of cosmology'', but currently the observational data
cannot tighten constrains on the physics at work at the very earliest
phase in its evolution. In this picture, shortly after the big bang,
13.8 billion years ago, the Universe was dominated by an energy field
with negative pressure that drove a period of accelerated expansion,
``the inflation'' phase. Since then the Universe has expanded, cooled
down, and changed from a radiation- to a matter-dominated
composition. If its content is dominated by a composition of baryonic
and cold dark matter one would expect a decelerated expansion of the
Universe. However by using type Ia supernovae (SNIa), as standard
candles, a surprising discovery has been made, that the expansion of
the Universe is undergoing a second epoch of acceleration (Riess et
al. 1998, Perlmutter et al. 1999, this research was awarded a the
Nobel prize in physics 2011). The reason for this recent accelerated
expansion is still a mystery and points to an additional phase of
negative pressure contribution of the mass-energy field and a possible
modification of Einstein's general relativity. Thus measuring the
recent acceleration will provide an additional route to probe the
equation of state and the interplay of dark energy.

\smallskip

\noindent Techniques to measure expansion rates in the Universe were 
already explored in 1962 by Sandage (Sandage 1962), but the
technological limitations at these time kept these measurements out of
reach.  It took more than 30 years for this idea to be revisited
and Loeb proposed 1998 to use Lyman-alpha forest absorption lines
toward quasars for this measurement. The author concluded that the
signal might be marginally detectable with a 10-m class telescope
(Loeb 1998). Now with the E-ELT, a 40-m class telescope, to be built
in the near future it seems possible to perform the ``Loeb' test''
with a specially designed spectrograph (for more information on the
CODEX\footnote{``Cosmic Dynamics Experiment''}-like experiments see
e.g.  Pasquini et al. 2005; Liske et al. 2008a, Liske et al. 2008b,
Maiolino et al. 2013). Unfortunately this test is greatly affected by
the cut-off restriction of ground based observations introduced by the
atmosphere. Due to this cut-off, Lyman-alpha photons can only trace
redshifts (z\,$\geq$\,1.65) at which most of the cosmological models
describe the expansion rate of the Universe by deceleration only.  SKA
observations in contrast, of the neutral hydrogen (HI) signals of Milky
Way-type galaxies up to redshifts of unity, are not greatly affected by
the atmosphere or ionosphere. This redshift regime is the one at which
most of the ``nearby'' acceleration takes place and therefore is the most
promising to investigate the influence of dark energy in the Universe.

\smallskip

\noindent The real time measurements of the cosmic acceleration its a very
appealing experiment that promises a model independent confirmation of
dark energy and a test to distinguish between cosmological
models. Despite the technical challenges this experiment may face, is
there a feasible SKA experiment in the radio regime that could measure
the frequency shift caused by the evolving Universe?

\begin{figure}
  \centering
  \includegraphics[width=13cm]{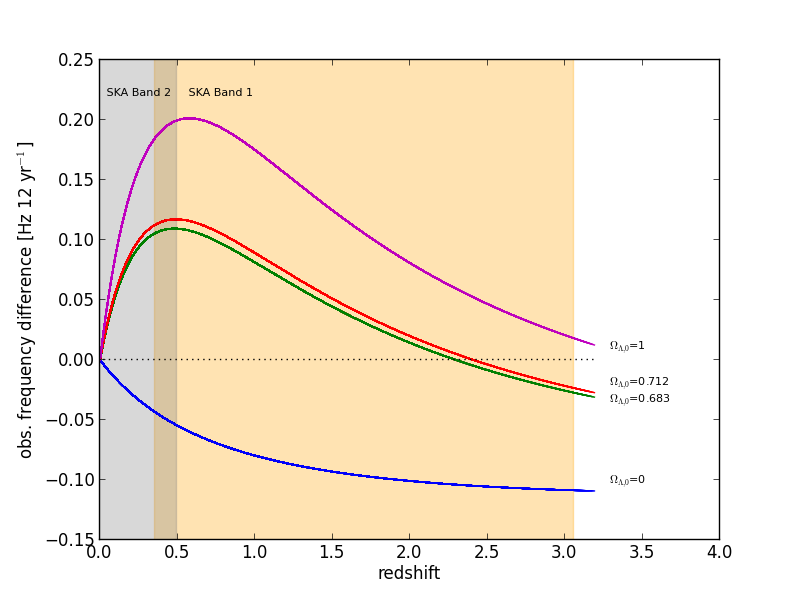}

  \caption{The expected redshift drift introduced by the cosmic
    acceleration against the redshift regime covered by the SKA. The
    redshift drift in 12\,years is shown as a change in frequency of
    the neutral hydrogen signal for various $\Lambda$CDM cosmologies
    ($\Omega_{m}$\,=\,0.27; $\Omega_{\Lambda}$\,=\,0.0, WMAP, Planck,
    1). The characteristic frequency shift shows a maximun shift of
    about 0.1~Hz, this translates into $\rm dz/dt \sim 10^{-10}$ or
    $\sim$3~cm in redshift- and velocity-space, respectively. The
    coloured background indicates the frequency regimes of band 1 and 2
    of the SKA basic design that are used to measure neutral hydrogen
    signals from galaxies (350 -- 1050 MHz, 950 MHz -- HI rest frame frequency).  }
  \label{FIG:dnudt}
\end{figure}

\section{The basic experiment}

\noindent The basic experiment is to detect the changes over time of
properties of individual galaxies caused by the expansion of the
Universe. Generally various observables could be used to constrain the
parameters needed to describe the expansion history of the Universe
which are the source brightness, the apparent angular size, the cosmic
parallax, or the redshift (see e.g. Gudmundsson \& Bj\"ornsson 2002,
Quercellini et al. 2012). However to measure the changes of the first
three observables seem to be out of reach to the current technical
capabilities of the SKA. But it seems feasible for the SKA to trace
the change of redshifts of individual galaxies and,  compared to the
CODEX-like experiment, a different approach is envisaged to measure
the cosmic acceleration. The basic experiment would make use of the
fast survey capabilities and the sensitivity to observe a billion of
HI galaxies up to redshift 1. Based on two HI surveys the task is to
combine the individual HI-line signals of a galaxy and statistically
merge up to a billion of these measurements. The high number
counts of galaxies will compensate the uncertainties of the
measurements and therefore permit a statistical detection of the
redshift drift (the initial setup of the experiment
has been described in Kl\"ockner 2012).\\

\noindent The redshift measurements are in general independent of a
cosmological model and rely only on the knowledge of the rest frame
frequency and the assumption that the fundamental constants do not
change over the evolution of the Universe (Kanekar 2012). In a
theoretical framework the change of redshifts can be described by
$\dot{z} = {\rm H(z)} - (1+z) {\rm H_0}$ and the fraction {$\rm
  H(z)/H_0$} is used to relate the acceleration to different kinds of
cosmological models. Therefore measuring the redshift drift at various
redshifts provides a unique test of different cosmologies (Balbi \&
Quercellini 2007). In the following a $\Lambda$CDM cosmology is
assumed and the Hubble parameter is described as

\begin{equation}
{\rm H(z)} = H_0  [{\rm \Omega}_{\rm m} \times (1 + z)^3 + {\rm
  \Omega}_{\rm \lambda} + {\rm \Omega}_{\rm k} (1 + z)^2]^{\frac{1}{2}},
\end{equation}

\noindent with ${\rm \Omega}_{\rm k} = 1 -( {\rm \Omega}_{\rm m} +
{\rm \Omega}_{\rm \lambda})$. Note that the first equation has been
written in such a way that the acceleration is positive ($\dot{\rm v}
> 0$) and the deceleration is negative ($\dot{\rm v} < 0$) defined in
the velocity frame. Figure~\ref{FIG:dnudt} displays the expected
frequency shift at different redshifts for various values of the cosmological
parameter, $\Omega_{\lambda}$, and a fixed $\Omega_m$ of
0.27. Depending on $\Omega_{\lambda}$ an acceleration of the Universe
is expected up to redshifts of 3 (positive frequency shift), after
this the expansion of the Cosmos will slow down (decelerate, negative
frequency shift). Furthermore, depending on $\Omega_{\lambda}$ the
frequency shift shows a distinct relationship with redshift and a
pronounced signature with a maximum at 0.4 and 0.6 in redshift. In the
case of the cosmological parameter measured by the WMAP and Planck
mission (Hinshaw et al. 2013, Planck Collaboration 2013) a maximum
frequency shift of 0.1~Hz can be expected after an observing period of
12~years.

\smallskip The direct measurement of the frequency shift and the
capability to distinguish between competing cosmological models
requires a precisions of up to a few percent (e.g. 1\% = 0.001~Hz). In
order to reach this kind of accuracy the main task of this experiment
is to utilise the HI line signals of a galaxy at 2 epochs and
statistically combine up to a billion of these measurements. In this
light the feasibility of this experiment depends on the projected
sensitivity estimates and the number counts of HI galaxies up to
cosmological redshifts. The expected number counts and the properties
of individual HI galaxies are based on the SAX-SKA sky simulation
(Obreschkow et al. 2009) and the image sensitivity can be determined
via the following equations (see e.g. Kl\"ockner et al. 2009):
\begin{equation}
{\rm \Delta I} = \frac{\rm SEFD}{   \eta_s \,  \sqrt{{\rm t} \, \Delta \nu  }},
\label{eq:sens}
\end{equation} 

\noindent with $\eta_s = 0.9$ is the system efficiency, t~integration
time [s], and $\Delta \nu$ is the channel width or bandwidth [Hz]. The
system equivalent flux density (SEFD) is determined via

\begin{equation}
{\rm SEFD} = \frac{ 2  {\rm k} {\rm T_{\rm sys}} }{ \rm A_{\rm eff}},
\end{equation}

\noindent where T$_{\rm sys}$ is the system temperature [K], k is the
Boltzmann constant, A$_{\rm eff}$ is the effective collecting area
[m$^2$]. \\

\noindent In order to investigate if the SKA is capable of detecting the
global signal of the redshift drift shown in Figure~\ref{FIG:dnudt} a
generic observational setup is assumed. The experiment would based on
two-``all-sky'' HI surveys with the following system parameters: SKA
(A$_{\rm eff}$ / T$_{\rm sys}$ = 13.000 m$^2$ K$^{-1}$), 1 hour
integration per pointing, 20 sq. degrees field of view, survey
coverage of 30.000 sq. degrees.  This setup will result in a
sensitivity of about 45~$\mu$Jy for the system channel width of
3.9~kHz and 28~mJy for 0.01~Hz wide channels (10\% accuracy). Based on
the sensitivity limit of 45~$\mu$Jy the expected number count of HI
galaxies in the redshift range z = 0.2-1 is of the order of $10^7$
sources ($\rm \# N$). To derive the frequency shift from the observed
HI line spectra of two epochs two approaches can be applied. Either
combining the cross-correlation spectrum of each source or fitting
signal components to line spectra and determine their statistical
means.

In the first case one would use the high-resolution spectra of the two
epochs and stack the power spectrum of the cross-correlation spectra
of all galaxies. Assuming that the noise properties of these spectra
are independent, the noise of the resulting averaged power-spectrum
will drop as $\rm \sim 1/\sqrt{\rm \# N}$ to 2.8~$\mu$Jy and
therefore allowing the detection of the global signal of the redshift
drift at a significance of $\sim 15 \sigma$.

In the second case one would use the low-resolution spectra and
determine the line properties by fitting a analytic function to the
line profile (e.g. using a busy function Westmeier et al. 2013). In
this way one assumes that all the observational parameters (the space
velocity vector of the observatory) can be modeled with such precision
that the residual data do not show any systematic effects above
millimeter accuracy. The difference of the modelled centre frequencies
determines the cosmic signal and its uncertainty will drop with $\rm
\sim 1/\sqrt{\rm \# N}$. Reversing the arguments, based on the
assumption that the uncertainties need to match some fraction of
cms$^{-1}$, this experiment is only possible if the redshifts or the
velocities of $10^7$ galaxies can be estimated at the observatory to a
precision of 10~ms$^{-1}$.

\noindent Based on the generic observational setup both methods
indicates that in general the SKA would be capable to
measure the global signal of the redshift drift.

\section{A feasible experiment with the SKA}

\noindent Tracing the signature of the cosmic acceleration at various
redshifts would be the ultimate experiment to test cosmological models.
In the case of a $\Lambda$CDM cosmology the redshift drift shown in
Figure~\ref{FIG:dnudt} indicates a characteristic feature up to
redshift of unity. The anticipated system performance of the SKA is an
ideal match to one of the key requirements of this experiment. In
particular the sensitivity figures of the SKA will be optimised to
trace the HI line (in emission) of Milky Way-like galaxies up to a
redshift of unity and therefore will produce the most complete
HI/redshift surveys in this redshift range.

\smallskip

\noindent The main aim of this experiment is to measure a shift in
frequency of about 0.1~Hz over a period of about 12~years (2.9
cms$^{-1}$ in velocity space and 10$^{-10}$ in redshift
space). However these measurements may suffer from various systematic effects
influencing the accuracy of the experiment. The potential
contributions to the uncertainties of the redshift drift can be
grouped as follows: (a)~The probe of the redshift drift at
cosmological distances show intrinsic variations. (b)~The model of the
Earth position in space and the observatory is insufficient for the
accuracy needed.  (c)~The technical hardware of the observatory.

Case~(a): Probing redshifts via the neutral hydrogen signal seems to
be one of the most promising probes in the radio regime, even if our
current knowledge of HI selected galaxies is limited up to redshifts
of 0.2. Generally these galaxies do not show any evidence for clustering or
a preference to populate over-dense regions in the Universe and hence
their redshift estimates are less affected than optically selected
galaxies (Papastergis et al. 2013; Kl\"ockner \& Romano-Diaz in
prep.). The contribution of the peculiar motion to the apparent
redshift has been estimated using the SAX-SKA- and the
millenium-simulations (Obreschkow et al. 2009, De Lucia \& Blaizot
2007).  For the individual galaxies, with redshifts larger than z=0.2,
the peculiar motion in redshift space has been estimated to $\rm
d(z+z_{pec})/dt \sim 10^{-14}$. This effect is a factor 10 smaller
than the cosmological signal at percent level and can be neglected in
the future.

Case~(b): The most challenging step in observing an extragalactic
redshift at the needed precision is to relate this measurement to an
accurate reference system in time and celestial direction
(e.g. Barycentre see Lindegren \& Dravins 2003). Furthermore, the
astrometry, the time standard, and the pointing accuracy can influence
the modelling of the line-of-sight doppler shift contribution of the
observatory. To match the precision requirements of the redshift-drift
estimates the pointing accuracy need to be of the order of
1~arcsec. The accuracy in astrometry and the distribution of the time
standard is related to the accuracy of the correlator model, including
an Earth model and the JPL ephemeris. However changes in the position
of the observatory e.g. due to tides (Solid-Earth: 30~cm, B. Campbell
private comm.), ocean- or atmospheric-loading (2~cm, B. Campbell
private comm.)  etc. need to be taken into account in the
post-processing of the full-sky survey.

Case~(c): The major technical challenge to overcome is the constraint
on the long time stability of the SKA system. The key is to reduce or
handle systematic effects of the observatory in timing precision and
frequency stability. Such requirements are difficult to evaluate and
need to be assessed in more detail in future
discussions. Nevertheless, it is assumed that pulsar observations will
be used to monitor the long time stability of the system. These
observations are sensitive enough to detect systematic effects within
the observatory like time drifts or global changes of the
observatory's 3-d velocity vector in the JPL ephemeris. Modelling of
the arrival time of the pulse of a network of pulsars should enable us
to correct for such systematics effects (Champion et al. 2010, Hobbs
et al. 2012).

An additional technical limitation is the number of correlator
channels in order to fully trace the neutral hydrogen signal of
galaxy. Generally the signal properties of the HI emission depends on
the rotational velocity and the inclination of the galaxy. Assuming
that the majority of the surveyed galaxies are Milky Way-type galaxies
with a rotational velocity of 300~kms$^{-1}$ a spectral window of
500~kms$^{-1}$ would be sufficient to trace the entire
signal. Tracing such a spectral window with the precision needed to
estimate the redshift drift $\sim$10$^{8-9}$ correlator channels are
needed. Such a high number of channels could be realised by a
dedicated pipeline, streaming the data from the predefined spectral
window into a software correlator. The setup of the spectral windows
and how the observations would be organised to set the frequency
standard needs to be investigated in the future.

\subsection{The experiments}

\noindent In order to investigate the feasibility of tracing the
signature of cosmic acceleration at various redshifts two approaches
with low- and high-spectral resolution datasets are discussed.

In case of low-spectral resolution datasets and an observing precision
of 10~ms$^{-1}$, the main constraint is to measure redshifts of $\sim
10^7$ galaxies per redshift interval.  In general these high number
counts can be achieved using an integration time of about 6 hours per
pointing, assuming the same observational setup as used in the
following discussion for the high-spectral-resolution case. The survey
time for this experiment would be of the order of one year with a 10\%
precision on the redshift drift measurement. In order to archive
higher precision the observing precision and hence the channel resolution
in frequency space needs to be adapted.

In the following the feasibility of this experiment is explored by
high-spectral resolution datasets ($\Delta \nu = 0.001, 0.002, 0.005,
0.01$~Hz). This discussion will address both phases of the SKA, the
SKA and a SKA$_1$ with hypothetically channel width of this order.

\smallskip

\begin{figure}
  \centering
  \includegraphics[width=6cm]{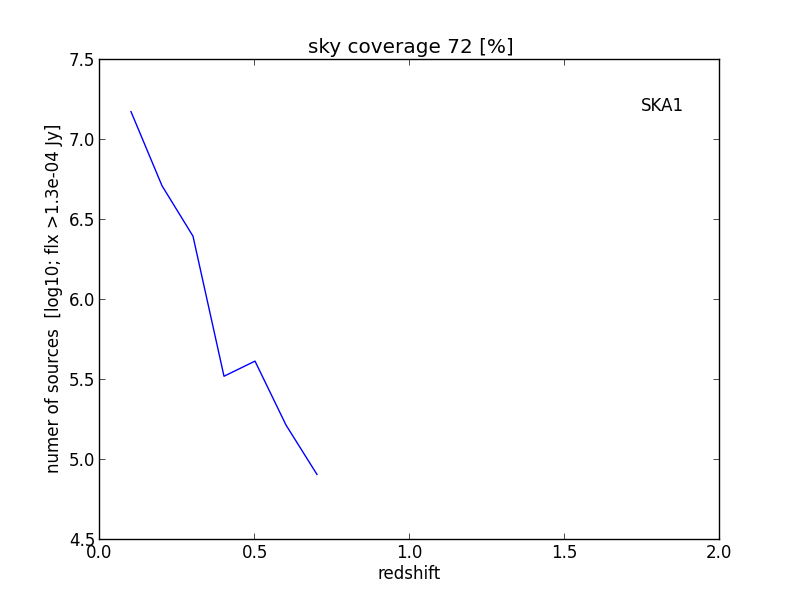}
  \includegraphics[width=6cm]{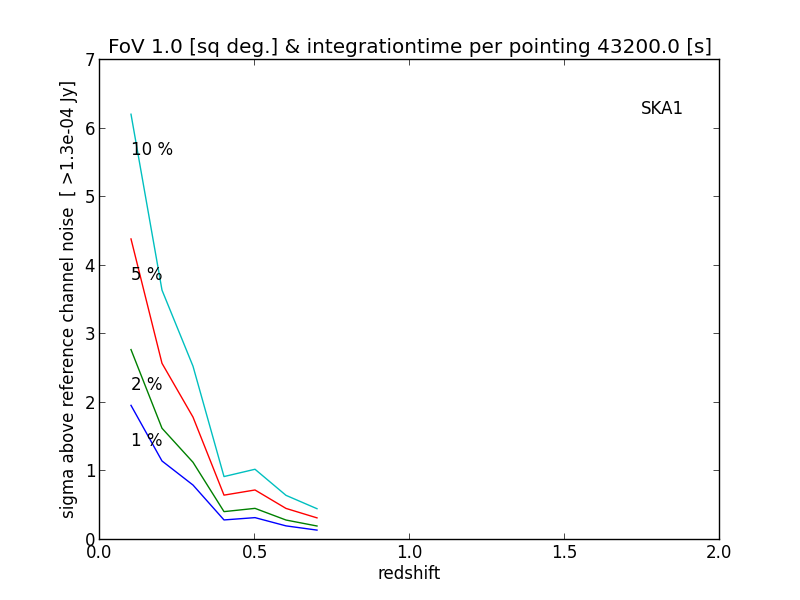}

  \includegraphics[width=6cm]{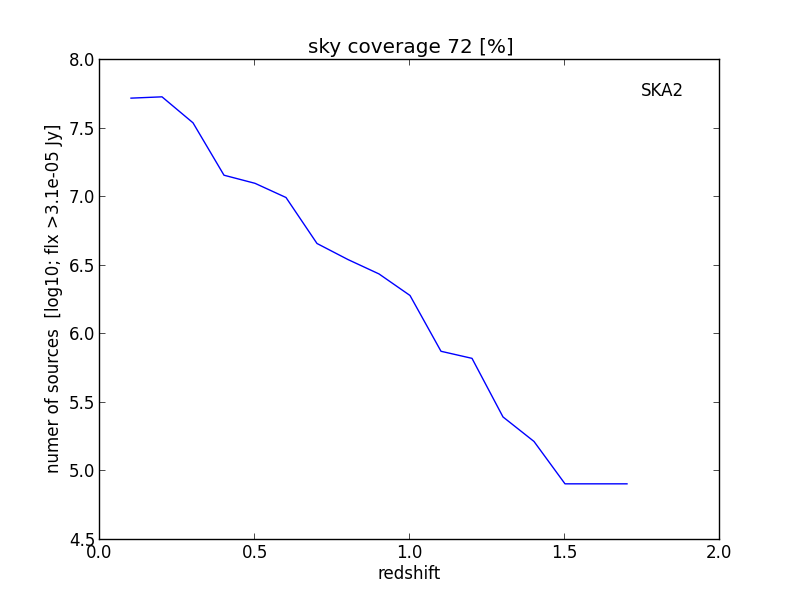}
  \includegraphics[width=6cm]{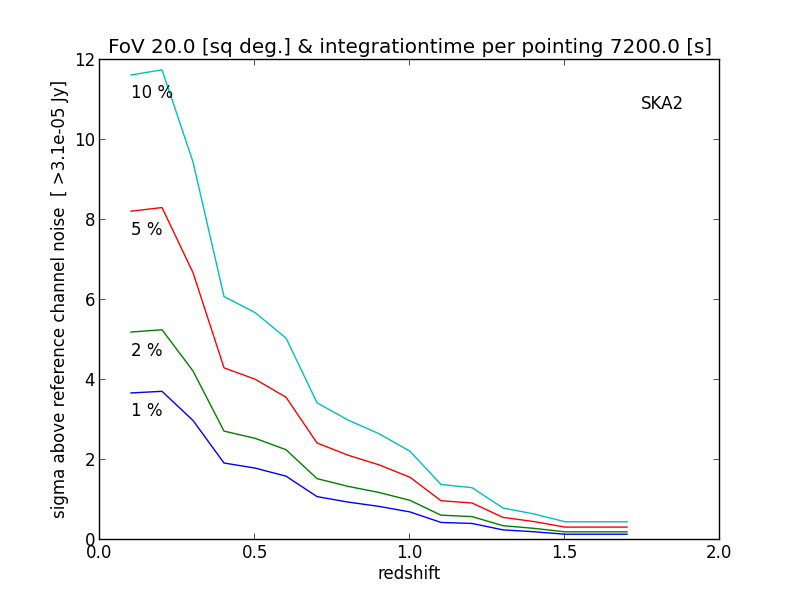}

  \caption{The basic ingredient of this experiment is the high number
    counts of galaxies and the sensitivity. The figures show the
    number counts and the significance per channel precision of
    detection versus redshift. The top panel show the results for a
    hypothetical SKA$_1$ with increased channel resolutions and the
    bottom displays the results for the full SKA.}
  \label{FIG:results}
\end{figure}

\bigskip

\noindent{\bf SKA Phase 1 (SKA$_1$)}

\smallskip

\noindent For the SKA$_1$ a generic observational setup is assumed
with the following system parameters: SKA$_1$ (A$_{\rm eff}$ / T$_{\rm
  sys}$ = 1300 m$^2$ K$^{-1}$), various channel resolutions, 12 hours
integration per pointing, 1 sq. degrees field of view, survey coverage
of 30000 sq.~degrees. The results are shown in the top panels of
Figure~\ref{FIG:results}. The results indicate that the number counts
would be sufficient to detect a redshift drift at redshifts less than
0.3 with 5\% to 10\% precision. These results are somewhat misleading
because to survey the entire sky with this setup would require
42~years which is not feasible, but it shows the importance of survey
speed to this experiment.

\smallskip 

\noindent In summary, the redshift drift experiment with the SKA$_1$
or 30\% of the SKA is not possible. Despite the fact that the
cosmological redshift drift can not be observed, a pathfinder
experiment could be initiated to aim for 10~ms$^{-1}$ accuracy to test the
``low spectral resolution'' case of the SKA$_2$ experiment. The
anticipated accuracy of 10~ms$^{-1}$ is already a factor 10 better
with respect to the radial velocity measurements of nearby standart
stars and may provide already some clues to the systematics and the
ephemeris (Chubak et al. 2012).

\newpage

\noindent{\bf SKA Phase 2 (SKA)}

\smallskip

\noindent For the SKA a generic observational setup is assumed with
the following system parameters: SKA (A$_{\rm eff}$ / T$_{\rm sys}$ =
13000~m$^2$ K$^{-1}$), various channel resolutions, 2 hour
integration per pointing, 20~sq.~degrees field of view, survey
coverage of 30000~sq.~degrees. The observations would take 125~days
and the results are shown in the bottom panel of
Figure~\ref{FIG:results} .  The results show that the number counts
would be sufficient enough to trace the functional dependency of the
frequency shift caused by the cosmological expansion up to redshift of
unity. The level of precision reached is a few percent and may even
reach the percent level if the integration time per pointing is 12~hours.

\noindent Due to the relatively short survey duration this experiment could even
be done several times within the life time of the SKA of 50~years.

\section{Discussion \& Summary}

\noindent Measuring the expansion history of the Universe includes
distances and the linear growth of density perturbations, and a
combination of both observed at different epochs. The observations at
different epochs allows for a direct measure of the expansion history,
whereas SNIa surveys, weak lensing (Heavens 2003) and Baryon Acoustic
Oscillations in the galaxy power spectrum (BAO; Wang 2006) are
generally considered to be indirect probes of the acceleration. Their
results rely on a priori knowledge of the cosmological model and even
simple parameterisations of dark energy properties can result in
misleading conclusions (e.g. Bassett et al. 2004, Shapiro \& Turner
2006). In this light redshift-drift measurements are direct probes and
rely only on the knowledge of the rest frame frequency of the measured
signal. Its uncertainty in testing cosmological models is the
knowledge of H$_0$. Compared to other probes with many more systematic
uncertainties in parameterisation and calibration (e.g. cosmic
chronometers; Moresco et al. 2012), the HI measurements offer less
biased measurement and can be assumed to be a model-independent
consistency test of cosmological theories. In addition, redshift drift
measurements are sensitive to different combinations of cosmological
parameters and thus combined with other probes can break degeneracies
and will place stronger constrains on cosmologies (e.g. CMB, BAO, weak
lensing). This complementarity nature of the redshift drift has been
shown e.g. for the E-ELT case improving the CMB constrain by a factor
of 2-3 and even has the potential to constrain new physics (Martinelli
et al. 2012, Vielzeuf \& Martins 2012).

The SKA will be able to measure the redshift drift to levels of a few
percent precision up to redshifts of unity. The quoted measurement
accuracy can be used to derive first constrains on cosmological model
and detailed forecasting on the accuracy of cosmological parameters
will be addressed in future investigations. However these precision
enables us to compare the SKA experiment already to a study of optical
redshift-drift measurements. This study indicates for a dynamical dark
energy wCDM cosmology the greatest leverage on figure of merit (FOM)
of the dark energy properties. In particular, an excess of 100 of FOM
for dark energy at 1\% precision and 1000, if combined with the
CMB. For a lower precision of 5\% the FOM is 290 if combined with the
CMB and the Hubble constant (Kim et al. 2015).

\bigskip

\noindent Real time cosmology is possible with the SKA. The SKA is the
only experiment that enables us to trace the nature of the ``close
by'' acceleration and may provide essential information on possible
mechanisms in place during the inflation phase and the evolution of
the Universe. However the proposed experiment does need the full
sensitivity of the SKA array, the survey speed to detect a billion
galaxies, and its technical requirements might imply some minor
changes of the SKA baseline design. In addition, this experiment would
benefit from a larger field of view of about 40~sq.~deg. per pointing
to allow for more sensitive observations per pointing. Nevertheless,
the already relatively short survey time of the SKA allows for a even
more ambitions experiment, measuring the cosmic jerk term. Such
measurement might be feasible if the redshift drift experiment is
performed several times within the life time of the SKA of 50 years.

\noindent Assuming the observational systematics can be modeled to
1$^{-3}$ms$^{-1}$ accuracy, the required spectral resolution of 10\%
(0.01~Hz) can be sufficiently reduced by a factor of hundred or
more. In case the correlator can not provide the spectral resolution
needed for this experiment, raw-data of individual galaxies of Band 1
or Band 2, or a part of it, could to be streamed to a dedicated
software-correlator pipeline. In both cases the SKA will be
able to detect the global signal of the redshift drift and with a
dedicated observational setup the SKA will measure the redshift
dependency of the redshift drift.

\noindent Even though this is a full SKA experiment; there are precursor
observations possible with the SKA in Phase 1 to investigate the
systematic effects within the ephemeris. These experiments would aim
to investigate the properties of the 3-dimensional velocity vector of
the Earth at resolutions of meter/s using the HI signal of nearby
galaxies and HI in absorption toward quasars. Measurements of the
3-dimensional velocity vector will not only have cosmological
applications they may also have the potential to model the Earth's
gravitational field and could open up new synergies between the SKA
and lunar ranging experiments.

\noindent Finally, the measurements of the redshift drift in redshift
space by the SKA and the E-ELT (both sampling different redshift
ranges) and the combination of their measurements are the only
experiments that will fully trace the real-time evolution of the dark
energy in the Universe.

\end{document}